\documentclass[12pt]{article}
\usepackage{graphicx}
\title{Glassy states in lattice models with
many coexisting crystalline phases}
\author{M. J. de Oliveira$^{(1)}$,
A. Petri$^{(2,3)}$ and T. Tom\'e$^{(1)}$\\
$^{(1)}$ Instituto de F\'{\i}sica, Universidade de
S\~ao Paulo, Caixa Postal 66318,\\
05315-970 S\~ao Paulo, S\~ao Paulo, Brazil \\
$^{(2)}$CNR, Istituto "O.M. Corbino",\\
via del Fosso del Cavaliere 100, 00133 Roma, Italy \\
$^{(3)}$Unit\`a INFM, Universit\`a  La Sapienza,\\ p.le Aldo Moro
2, 00185 Roma, Italy}

\begin{document}
\maketitle

\begin{abstract}
We study the emergence of glassy states after a sudden cooling in
lattice models with short range interactions and without any a
priori quenched disorder. The glassy state emerges whenever the
equilibrium model possesses a sufficient number of coexisting
crystalline phases at low temperatures, provided the thermodynamic
limit be taken before the infinite time limit. This result is
obtained through simulations of the time relaxation of the
standard Potts model and some exclusion models equipped with a
local stochastic dynamics on a square lattice.
\end{abstract}

\section{Introduction}

Recent years have seen a growing interest in models that,
albeit possessing ordered ground states, are capable
of displaying  glassy  behaviour
\cite{lipowski1,swift,franz01,biroli02,weigt02,pica02}.
Other models
contain a priori ingredients that
prevent them from reaching or even possessing a crystalline state;
examples of these ingredients are: quenched
disorder \cite{young98}, competing interactions \cite{young98},
polydispersity \cite{santen00}, modified long-range potentials
\cite{angelani01}, and constrained interactions \cite{kob93,davison01}.
While these factors may fit naturally in the description of
systems characterized by a quenched disorder, like spin glasses, they
seem less suitable to the
description  of  structural glasses since
they explicitly suppress crystallization, unlike most of real
systems that, on the contrary, are potentially good crystal formers.
The use of models with such a priori ingredients is motivated by
both practical and theoretical reasons.
From a practical point of view, numerical simulations are
limited to relatively small lattices, so that large
collective effects that in real systems could be at the origin of
glassy behavior can be very depressed: e. g.  molecular dynamics
simulation of  identical atoms interacting via a Lennard-Jones
potential eventually  produces crystalline configurations also
in the presence of very fast cooling.
From a theoretical point of view,  thermodynamic stability
excludes the possibility
that systems with translational invariance and short
range interactions  could possess metastable states at finite
temperatures \cite{griffiths64,yeomans92}.
On the other hand, formation of large and
pure crystals is uncommon in nature, and even their growth in a
laboratory is not a simple task,
despite  the fact that  freezing of matter into non-crystalline phases
is contrary to what would be expected from equilibrium thermodynamics.

For the  above reasons  it would be  desirable to have models
capable of displaying glassy behavior  without the  introduction
of any kind of quenched disorder or special interactions, and to
have simple models sheding some light on the mechanisms that drive
crystal formers away from crystallization.  The aim of the present
letter is to show under which circumstance some simple lattice
models, with neither quenched disorder nor constraining rules, are
capable of relaxing into a glassy state. We will specifically
investigate the Potts and the exclusion models, showing that the
conditions which prevent the ordering are the same in the two
models.

\section{Glassy states}

It has been observed that some exclusion models
\cite{evans93,privman00,runnels}, where crystalline states are
expected as a consequence of the given dynamical evolution rules,
(see sections below) are pinned into some noncrystalline state. In
these models the occupation of a lattice site by a particle
excludes the occupation of a certain set of neighboring sites by
other particles and it may happen that the effects of excluded
volume prevent the system from settling down to the ordered state
\cite{privman00}. It has been found in particular
\cite{wang93a,wang93b,eisenberg98} that when on a square lattice
the exclusion of first and second nearest-neighbours is imposed,
the system slowly approaches a state in which particles are
arranged to form different macroscopic crystalline domains.
However, when exclusion of third nearest-neighbours is also
imposed, the system jams into noncrystalline configurations
\cite{eisenberg00}.

The properties of the above models suggest that
the glassy state (in the sense quantitatively
defined in the next sections) emerges after a sudden
cooling in  systems that,  when in thermodynamic equilibrium,
display a sufficient number of equivalent crystalline states
or coexisting ordered phases at low temperatures, and a
disordered phase at high temperatures. This is the case of the
Potts model \cite{wu82} and
exclusion models \cite{evans93,runnels} studied here.
When in equilibrium, they suffer a phase transition
which is continuous
if the number of ground states is small and  first-order
if it is large. We implement the time relaxation of these models
due to the cooling process by supplying them
with a local stochastic dynamics. By means of extensive
numerical simulations we find that whenever the equilibrium
model exhibits a continuous phase transition the stochastic system
relaxes towards the equilibrium order state, either a
crystalline or a polycrystalline. However, when the transition
is first-order, the stochastic system relaxes towards a
nonequilibrium glassy
state provided the thermodynamic limit be taken {\it before}
the infinite time limit. Notice the reversion in the order
of the limits when compared with the equilibrium case in which
the thermodynamic limit is taken after the infinite time limit.

\section{The Potts model}

In the $q$-state Potts model \cite{wu82}
each site of a regular lattice assumes one out of
$q$ different possible states, or colour. If two
nearest neighbour sites have the same colour their
energy contribution is zero, otherwise they contribute by an amount
$\epsilon >0$ to the system energy:
\begin{equation}
\label{e:1}
E=\sum_{(ij)} \epsilon(1-\delta_{\eta_i \eta_j}),
\end{equation}
where $\eta_i=1,2, \dots q$ and the sum is over the nearest neighbor
sites. There are $q$ degenerate ground states, with zero energy, each one
corresponding to having all sites the same colour.
We stress that the energy is proportional to the length $\ell$
of the line separating regions of distinct colours
(line of defects), $E=\epsilon \ell$.
We distinguish the states according to the value of $\ell$.
If $\ell=0$, there is no defect and the state is crystalline.
If $\ell>0$ but negligible, the state is polycrystalline,
otherwise, if $\ell$ is not neglible the state is a noncrystalline
state which we call a glassy state. By negligible we mean that the
density of defects $\rho_{\sf x}=\ell/N$ vanishes in
the thermodynamic limit.

The sudden cooling process is performed by simulating the Potts
model at zero temperature according to the Metropolis dynamics
\cite{metropolis53} starting with a random and uncorrelated
configuration to mimic the infinite temperature state. We start by
using Periodic Boundary Conditions (PBC). For long times, the
system is expected to relax towards the state of minimum energy,
$E=0$, where all sites have the same colour. However, this may not
happen and the system may be trapped in a noncrystalline
configuration, similar to Mondrian or van Doesburg paintings, with
many local traps with minimum energy in the shape of a
``\textsf{T}'' separating different domains.
This  behavior is well
known \cite{anderson89,derrida96} and is not found when Fixed
Boundary Conditions (FBC) are imposed to the lattice, i.e., when
one predetermined arbitrary colour  is assigned to the sites
beloging to the lattice boundary. With this prescription the
system is forced to relax to the crystalline state and, after a
slow relaxation which occur for an intermediate time regime, the
system eventually enters into a fast relaxation stage by which all
sites finally assume the same colour of the boundaries.

\begin{figure}
\begin{center}
\includegraphics*[scale=0.3]{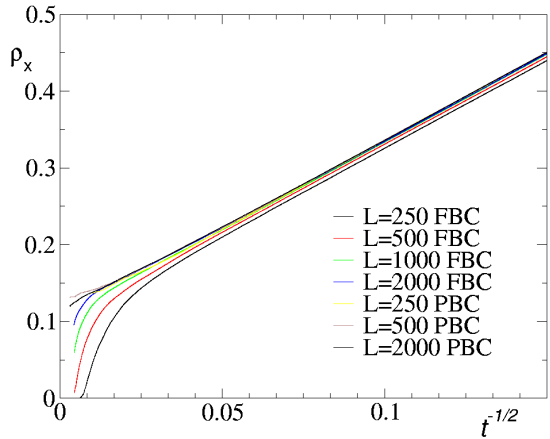}
\caption{Density of defects $\rho_{\sf x}$ versus $1/\sqrt t$ at
large times, for the $7$-state Potts model on a square lattice for
several linear sizes $L$ and different boundary conditions (PBC or
FBC).} \label{rf.1}
\end{center}
\end{figure}

\begin{figure}
\begin{center}
\includegraphics*[scale=0.3]{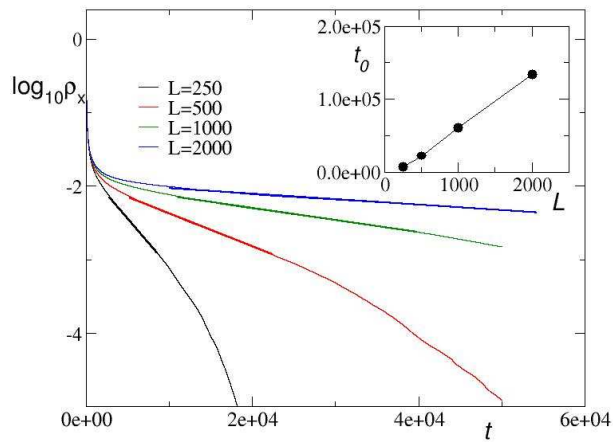}
\caption{Late stage relaxation to the ground state energy for 
the Potts model
with $q=7$ and FBC. An exponential regime
can be detected (thick dashed lines) whose characteristic time 
$t_0$ increases linearly  with
the lattice size (inset).} \label{rf.2}
\end{center}
\end{figure}

In fig.~\ref{rf.1} we have plotted the density of defects
$\rho_{\sf x}$ against $1/\sqrt t$ obtained from simulations of
the 7-state Potts model with both PBC and FBC on a square lattice
with $N=L\times L$ sites.
Up to an intermediate time regime all curves display the same
linear behavior, showing that the density of defects decays like
$1/\sqrt t$ \cite{bray94} independently of both the boundary
conditions and the lattice size. At long times the behavior is
distinct for each boundary condition. When PBC are used, the
system reaches configurations with a nonzero density of defects
$\rho_{\sf x}$.
 When FBC are
used, a fast (exponential) relaxation enters into action at a
characteristic time, the system eventually reaches the ground
state and $\rho_{\sf x}$ vanishes, as shown in fig. \ref{rf.2}. The
characteristic time increases with $L$ (inset of the figure). 
Extrapolation to the thermodynamic limit implies its divergence and 
the disappearance of
the final relaxation towards the ground state. According to
fig.~\ref{rf.1}, the curves corresponding to PBC and FBC approach
each other and are expected to become identical in this limit. The
extrapolation of the linear region in fig.~\ref{rf.1} gives a
nonzero value of the density of defects $\rho_{\sf x}^*$ when $t
\rightarrow \infty$. This nonzero residual density of defects
shows that the defects are not negligible in the thermodynamic
limit and we are faced with a glassy states.

\begin{figure}
\begin{center}
\includegraphics*[scale=0.2]{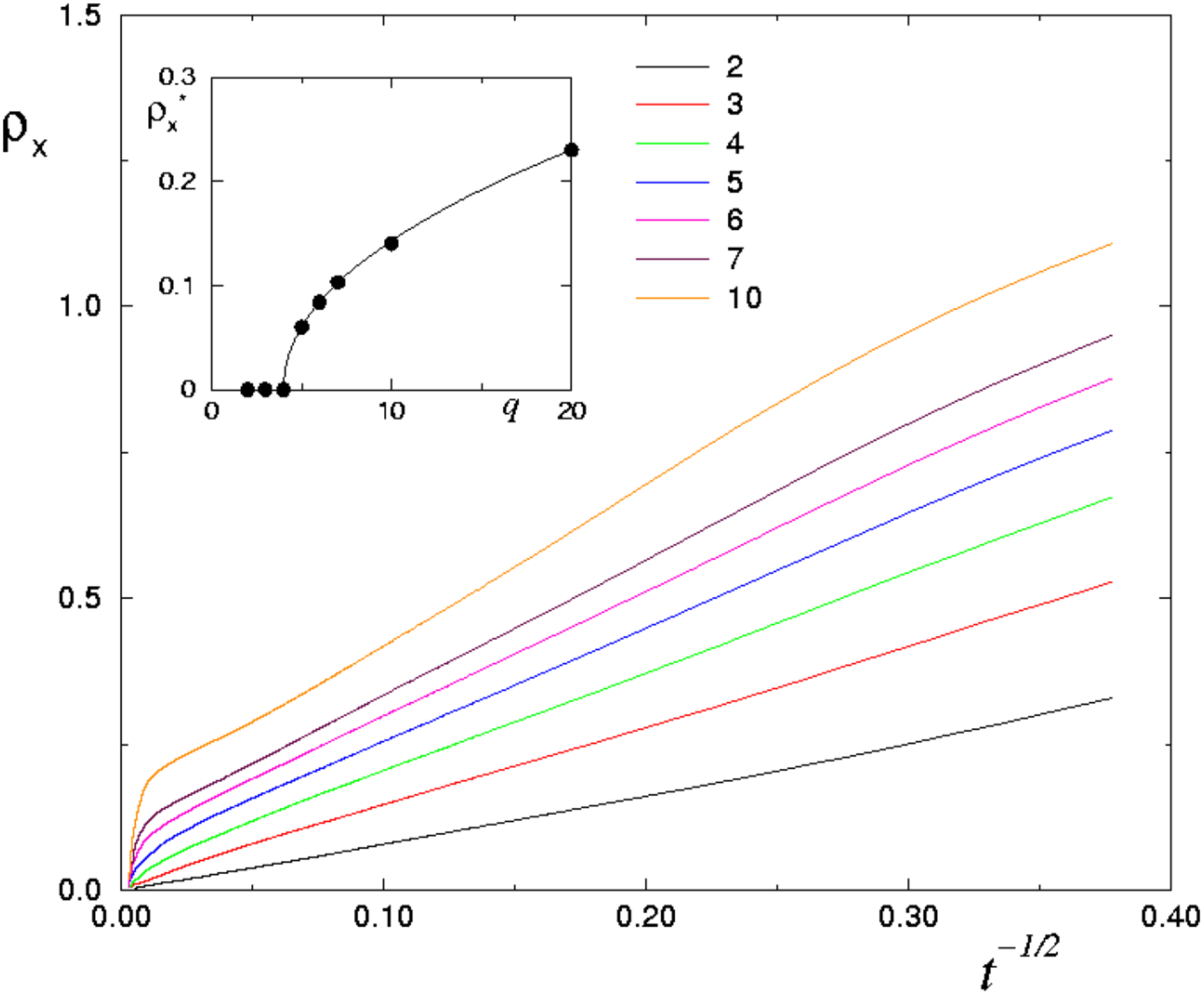}
\caption{ Density of defects $\rho_{\sf x}$ versus $1/\sqrt t$ for
the $q$-state Potts model with $q=2$ up to $q=10$ on a square
lattice of linear size  $L=10^3$ with FBC. The inset shows the
residual density of defects $\rho_{\sf x}^*$ versus $q$, including $q=20$. 
The continuous line is a fitting of the type $\rho_{\sf x}^* = A
(q-4)^\beta$. The best fitting gives $A=0.060$ and $\beta=0.49$.}
\label{rf.3}
\end{center}
\end{figure}

We have investigated the density of defects for several values of
$q$, as shown in fig.~\ref{rf.3}, and found that whenever $q>q_c=4$
the residual density of defects $\rho_{\sf x}^*$ is nonzero. In
the special case of $q = q_c$, there are logarithmic corrections
in the time decay and our results for $q=4$ suggest the behavior
$\rho_{\sf x} \approx (\ln t/t)^{1/2}$. For different values of
$q$, the residual densities of defects resulting from our analysis
are reported in the inset of fig.~\ref{rf.3}.

On the basis of the
results obtained,  we can draw the following conclusions:
{\it $i)$ for $q = 2$
the system relaxes towards a pure crystal (no defects);
$ii)$ for $q = 3,4$ the system forms a polycrystal (negligible
number of defects);
$iii)$ for $q > 4$ the system relaxes towards a
glassy state (nonzero density of defects). }

Existence of
low temperature noncrystalline states was predicted on a general
basis by Lifshitz \cite{lifshitz62} who argued that a
$d$-dimensional system quenched below its critical point does not
necessarily equilibrate into a pure state, in the presence of more
than $d+1$ ordered ground states. The competition among  different
possible ordering processes would drive the system away from
crystallization, although nothing is given to know about the length  
of the lines of defects.
An interesting investigation on the formation of 
ordered vs disorered structures in the quench of some two 
dimensional lattice gas models can be found in \cite{sadiq84}. 

\section{The exclusion models}

In exclusion models, adsorbed particles diffuse over the sites of
a regular lattice in such a way that each adsorbed particle
excludes the sites of a certain neighborhood from being occupied
by other particles.  The particles do not interact except through
excluded volume repulsion.

In order to give a quantitative definition of a
glassy state it can be considered tha case in which such a system
is placed in contact with a particle reservoir. The energy of the
system will be $-\mu n$ where $\mu>0$ is the chemical potential
and $n$ is the number of adsorbed particles. The ground state is
that of the closest packing because the number of deposited
particles is maximum. Here we set the zero of energy as that of
the ground state so that the energy is $E=\mu (n_{\rm cp}-n)$
where $n_{\rm cp}$ is the number of particles in the close-packed
state. With this definition, the energy is proportional to the
number of vacancies $n_{\rm x}=n_{\rm cp}-n$. Again we may define
a glassy state as being the state in which the density of
vacancies $\rho_{\rm x}=n_{\rm x}/N$ is nonzero in the
thermodynamic limit. Otherwise the state is either crystalline or
polycrystalline.

Sudden cooling process at zero temperature in these systems 
has been investigated in the literature (a recent review can be
found in \cite{privman00}). It can be performed
by simulating the model
according to the Metropolis dynamics starting with
an empty lattice. According to this dynamics
particles either diffuse or are adsorbed. In the absence of
diffusion a jamming state would be reached, in which all
lattice sites would be blocked \cite{evans93,privman00}. However diffusion of adsorbed
particles over the lattice allows for deposition of other
particles whenever a site previously blocked is made free.
In principle, a close-packed state may eventually be reached in which
particles are arranged into a regular periodic pattern.

We have considered exclusion models on a square lattice with
different number of ground states.  The first model studied ($N_1$
model) has two ground states and is defined by imposing the
exclusion of the nearest-neighbour sites. This corresponds to a
system of hard squares, each one having an area equivalent to 2
unit cells and thus covering 2 sites of the lattice. The
close-packed state in this case consists of a checkerboard-like
arangements of particles and can occur on two different
sub-lattices. Models with a different number of ground states can
be designed in a similar way, by the exclusion of a suitable set
of neighbours.

It is known \cite{wang93a} that for the  $N_1$ model,  the
deposition-diffusion process leads to ordering into one of the two
ground states. For determining the behavior of models with a
different number of ground states we have investigated numerically
the models with exclusion of both first and second neighbors,
$N_2$ \cite{wang93b,eisenberg98} and of third neighbors as well,
$N_3$ \cite{eisenberg00}. The $N_2$ model corresponds to a system
of hard plane objects, each one covering 4 sites of the lattice;
whereas in $N_3$ each hard object covers 5 sites. In analogy with
the case of the Potts model we have plotted the density of
vacancies $\rho_{\sf x}$ versus  $1/\sqrt t$.

\begin{figure}
\begin{center}
\includegraphics*[scale=0.4]{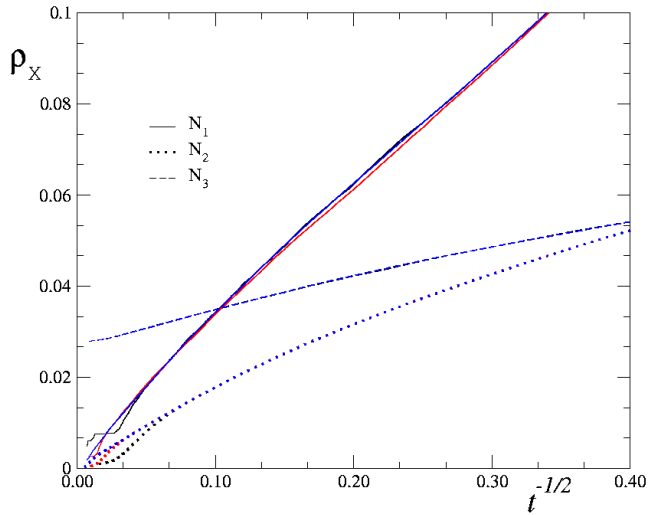}
\caption{Density of vacancies $\rho_{\rm x}$ versus $1/\sqrt t$
for the exclusion models $N_1$, $N_2$ and $N_3$ on a square
lattice. For each model, the linear size of the lattice are
$L=50$, $100$ and $200$.} \label{rf.4}
\end{center}
\end{figure}

Fig.~\ref{rf.4} shows the decay of $\rho_{\rm x}$ for the $N_3$
model. Extrapolation  to $L=\infty$ allows to conclude that in
this case $\rho_{\rm x} \rightarrow \rho_{\rm x}^*>0$ when
$t\rightarrow \infty$. On the contrary, simulations of the $N_2$
model reported also in fig.~\ref{rf.4} show that in this case
$\rho_{\rm x} \rightarrow 0$ when $t\rightarrow \infty$, although
as $\rho_{\rm x} \approx (\ln t/ t)^{1/2}$. By considering the
number $q$ of sites covered per hard plane object in each model we
are thus lead to the same conclusions as for the Potts model: {\it
$i)$ in the $N_1$ model, $q=2$ and the system relaxes towards a
pure crystal (no vacancies); $ii)$ in the $N_2$ model, $q=4$ and
the system forms a polycrystal (negligible number of vacancies);
$iii)$ in the $N_3$ model, $q=5$ and the system relaxes towards a
glassy state (nonzero density of vacancies).}

\section{Discussion}

These conclusions show that the ability of these systems to form
 glassy states depends on the number of equivalent crystalline
states as a consequence of the competition among
different ordered domains.
The critical number $q_c$ of equivalent crystalline states
above which the glassy state emerges is found to be $q_c=4$ both
for the Potts and the exclusion models on a square lattice.

We have also perfomed numerical simulations of the Potts
model on a triangular lattice for values of $q$ ranging
from $q=2$ up to $q=12$ and found that the density of
defects vanishes in the infinite time limit.
These results are consistent with the results obtained
by Sahni et al \cite{sahni83} in which domains grow
without limits in the Potts model on a triangular
lattice but are pinned on a square lattice.

The results obtained on the square lattice suggest that the
observed glassy states might be related to the presence of
a first-order transition.
In fact, the equilibrium two-dimensional Potts model suffers a phase
transition from a disordered state, at high temperatures,
to an ordered state with $q$ coexisting phases, at low
temperatures, which is continuous for $q\le4$, and
first-order for $q>4$ \cite{wu82}.
Analogous features have been observed in the
exclusion models studied here:
In equilibrium,  $N_1$ and $N_2$ models display
a second-order transition
whereas the $N_3$ model show a first-order transition
\cite{runnels,orban82,domany77}.

\section{Conclusion}

The results presented in this letter show that
systems with translational invariance and short range interactions
may be driven away from crystallization by the competition among a number
of coexisting ordered states. For the models considered here, i. e.
the Potts model
and the exclusion models on the square lattice, the number of states
above which ordering is destroyed corresponds to the appearance of
a first-order transition in
the equilibrium diagram of the systems and is the same for
the two models.

These results are obtained  by reversing the usual
order in which limits are taken when doing equilibrium statistical
mechanics, i. e. the limit $L\rightarrow\infty$ is taken before
$t\rightarrow \infty$. Within this approach, our simulations indicates that
the characteristic time for the system to relax to the ground states
may grow faster than the system size, in the sense that thermodynamic
equilibrium is never attained \cite{schulmann,biroli01}.


\begin{thebibliography}{0}

\bibitem{lipowski1}
Lipowski A.,
J. Phys. {\bf A 30} (1997), 7365.

\bibitem{swift}
Swift M. R.,  Bokil H., Travasso R. D. M. \and Bray A. J.,
Phys. Rev. {\bf B 62} (2000), 11494.

\bibitem{franz01}
Franz, S,  Mezard M., Ricci-Tersenghi F., Weigt M.
\and Zecchina R.,
Europhys. Lett. {\bf 55} (2001), 465.

\bibitem{biroli02}
Biroli G. \and  Mezard M.,
Phys. Rev. Lett. {\bf 88} (2002), 25501.

\bibitem{weigt02}
Weigt M. \and  Hartmann A. K.,
cond-mat/0210054.

\bibitem{pica02}
Pica Ciamarra M., Tarzia, M., de Candia A. \and Coniglio A.,
cond-mat/0210144.

\bibitem{young98}
A. P. Young Ed.,
{\em Spin Glasses and Random Fields}
World Scientific, Singapore
(1998).

\bibitem{santen00}
Santen L. \and Krauth W.,
Nature {\bf 405} (2000), 550.

\bibitem{angelani01}
Angelani L., Parisi G., Ruocco G. \and Viliani G.,
Phys. Rev. Lett. {\bf 87} (2001), 5502.

\bibitem{kob93}
Kob W. \and  Andersen H. C.,
Phys. Rev. {\bf E 48} (1993), 4364.

\bibitem{davison01}
Davison L., Sherrington D., Garahan J. P.  \and Buhot A.,
J. Phys. {\bf A 34} (2001), 5147.

\bibitem{griffiths64}
Griffiths R. B.,
J. Math. Phys.{\bf 5} (1964), 1215.

\bibitem{yeomans92}
Yeomans J. M., {\em  Statistical Mechanics of Phase
Transitions},
Oxford University Press, New York,
(1992).

\bibitem{tamarit}
Gleiser P. M., Tamarit  F. A. \and  Cannas S. A.,
Physica {\bf D 168-169} (2002), 73.

\bibitem{ruffo}
Antoni M. \and Ruffo S.,
Phys. Rev. {\bf E 52} (1995), 2361.

\bibitem{evans93}
Evans J. W.,
Rev. Mod. Phys. {\bf 65} (1993), 1281.

\bibitem{privman00}
Privman V., Colloids Surf. {\bf A 165} (2000), 231.

\bibitem{runnels}
Runnels L.K.,
{\em Phase Transitions and Critical Phenomena}
C. Domb \and M. S. Green Eds.,
Vol. {\bf 2}
Academic Press, London
(1972), p. 305.

\bibitem{wang93a}
Wang J. S., Nielaba P. \and Privman V.,
Physica {\bf A 199} (1993), 527.

\bibitem{wang93b}
Wang J. S., Nielaba P. \and Privman V.,
Mod. Phys. Lett. {\bf 7} (1993), 189.

\bibitem{eisenberg98}
Eisenberg E. \and  Baram A.,
Europhys. Lett. {\bf 44} (1998), 168.

\bibitem{eisenberg00}
Eisenberg E. \and  Baram A.,
J. Phys. {\bf A 33} (2000), 1729.

\bibitem{wu82}
Wu F. Y.,
Rev. Mod. Phys. {\bf 54} (1982), 235.

\bibitem{metropolis53}
Metropolis N., Rosenbluth A. W., Rosenbluth M. N.,
Teller A. H. \and  Teller E.,
J. Chem. Phys. {\bf 21} (1953), 1087.

\bibitem{anderson89}
Anderson, M. P.  Grest  G. S. \and  Srolovitz D. J.,
Phil. Mag. {\bf B 59} (1989), 293, and refs. therein.

\bibitem{derrida96}
Derrida B., de Oliveira  P. M. C. \and  Stauffer D.,
Physica {\bf A 224} (1996), 604.

\bibitem{bray94}
Bray A. J.,
Adv. Phys. {\bf 43} (1994), 357.

\bibitem{lifshitz62}
Lifshitz I. M.
Zh. Eksp. Teor. Fiz. {\bf 42} (1962), 1354,
transl. Sov. Phys. JETP {\bf 15} (1962), 939.

\bibitem{sadiq84}
Sadiq A. \and Binder K.,
J. Stat. Phys. {\bf 35} (1984), 517.

\bibitem{sahni83}
Sahni P. S.,  Srolovitz D. J.,  Grest G. S.
Anderson M. P.  \and  Safran  S. A.,
Phys. Rev. {\bf B  28} (1983), 2705.

\bibitem{orban82}
Orban J. \and Van Belle J. D.,
J. Phys. {\bf A 15} (1982), L501.

\bibitem{domany77}
Domany E., Schick M. \and  Walker J. S.,
Phys. Rev. Lett. {\bf 38} (1977), 1148.

\bibitem{schulmann}
Schulman L. S.,
{\em Finite Size Scaling and Numerical Simulations of Statistical Systems}
V. Privman  Ed.,
World Scientific, Singapore
(1990),
p. 489.

\bibitem{biroli01}
Biroli  G. and  Kurchan J.,
Physical Review {\bf E 64} (2001), 16101.

\end{thebibliography}
\end{document}